\documentclass[onecolumn]{IEEEtran}

\usepackage{amssymb,amsthm}
\usepackage[cmex10]{amsmath}
\usepackage{cite}

\newcommand{\ep}{\epsilon}
\newcommand{\diff}[1]{d#1}
\newcommand{\E}[1]{\bar{#1}}
\newcommand{\delt}[2]{\delta^{(#1,#2)}}
\newcommand{\divi}[2]{\frac{d#1}{d#2}}
\newcommand{\rdivi}[2]{\frac{\partial #1}{\partial #2}}
\newcommand{\xbar}{\tilde{x}}
\newcommand{\epbar}{\tilde{\ep}}
\newcommand{\calL}{\mathcal{L}}
\newcommand{\calR}{\mathcal{R}}
\newcommand{\rmax}{d_{c}}

\newtheorem{theorem}{Theorem}
\newtheorem{lemma}{Lemma}
\newtheorem{corollary}{Corollary}
\newtheorem{remark}{Remark}
\begin{document}
\title{Analytical Solution of Covariance Evolution \\
for Irregular LDPC Codes}

\author{Takayuki~Nozaki,~\IEEEmembership{Student~Member,~IEEE,}
        Kenta~Kasai,~\IEEEmembership{Member,~IEEE,}
        and~Kohichi~Sakaniwa,~\IEEEmembership{Member,~IEEE}%
\thanks{The authors are with 
Department of Communication and Integrated Systems, 
Tokyo Institute of Technology,
Tokyo 152-8550, Japan 
e-mail: \{nozaki, kenta, sakaniwa\}@comm.ss.titech.ac.jp .}
\thanks{Manuscript received January 1, 20**; revised January 1, 20**.
The material in this paper was presented in part 
at IEEE International Symposium on Information Theory (ISIT'09), 
Seoul, Korea, July 2009.}}

\markboth{IEEE Transactions on information theory,~Vol.~*, No.~*, January~20**}%
{Nozaki \MakeLowercase{\textit{et al.}}: Analytical solution of
covariance evolution for irregular LDPC codes}

\maketitle

\begin{abstract}
A scaling law developed by Amraoui et al.\
is a powerful technique to estimate the block error probability
of finite length low-density parity-check (LDPC) codes.
Solving a system of differential equations called covariance evolution
is a method to obtain the scaling parameter.
However, the covariance evolution has not been analytically solved.
In this paper, 
we present the analytical solution of the covariance evolution
for irregular LDPC code ensembles.
\end{abstract}

\IEEEpeerreviewmaketitle

\section{Introduction}
Gallager invented low-density parity-check (LDPC) codes \cite{Gallager}.
LDPC codes are linear codes defined by sparse bipartite graphs,
called \textit{Tanner graphs}.
\textit{Peeling algorithm} (PA) \cite{Luby97,mct} introduced by Luby et al.\
is a sequential iterative decoding algorithm 
for the binary erasure channel (BEC).
As PA proceeds, edges and nodes are progressively removed
from the original Tanner graph and the so-called
\textit{residual graph} is left at each iteration.
The residual graph at each iteration consists of 
variable nodes that are still unknown
and the check nodes and the edges connecting to those variable nodes.
The decoding successfully halts
if and only if the residual graph vanishes.
It is known that PA and brief propagation (BP) decoder
have the same decoding result.

The \textit{scaling law} developed by Amraoui et al.\ \cite{Amraoui09}
is a powerful technique to estimate the block and bit error probability
of finite length LDPC codes.
Let $r_i$ and $l_j$ be random variables representing
the number of edges connecting to the check nodes of degree $i$
and the variable nodes of degree $j$, respectively, in the residual graph.
Then, the \textit{scaling parameter} is obtained 
from the mean and the variance of $r_1$.
The means of $r_i$ and $l_j$ 
are determined from a system of differential equations
which was derived and analytically solved 
by Luby et al.\ \cite{Luby97}.
The covariances of $r_i$ and $l_j$
also satisfy a system of differential equations
called \textit{covariance evolution}
which was derived by Amraoui et al.\ \cite{Amraoui09}.
However, the analytical solution of the covariance evolution
has not been known.
Therefore, one had to resort to numerical computation to 
solve the covariance evolution.

In \cite{Amraoui07}, Amraoui et al.\ proposed 
an alternative way to determine the variance of $r_1$,
though only at the decoding threshold.
Thereby they have given the analytic expression for
the scaling parameters without using covariance evolution.
They used BP decoding instead of PA.
This method was applied to irregular repeat-accumulate codes in 
\cite{Pfister05,Andriyanova08}
and to turbo-like codes in \cite{Andriyanova09}
and was extended to binary memoryless symmetric channels in \cite{ezri2008}.

Denote by $\xi$ the total number of edges in the Tanner graph.
Let $\mu_i$ be the random variable 
which is 1 if the edge $i$ conveys an erasure message from a variable node
to a check node, and 0 otherwise, in the BP decoding.
The method in \cite{Amraoui07} analyzed the random variable
$M:=\sum_{i=1}^{\xi}\mu_i$ in the BP decoding and derived
the analytical expression for the variance of $M$.
Finally, they did make an unproved assumption that
the random variable $r_1 - E[r_1]$ in PA
is proportional to the random variable $M - E[M]$ in BP
and under this assumption they have given
the analytical solution for the variance of $r_1$.

However, no such assumption is needed
if the covariance evolution is solved analytically.
Moreover, we can obtain the variance of $r_1$ 
at any channel erasure probability.
In this paper,
we present the analytical solution of the covariance evolution
for irregular LDPC code ensembles.
\section{Preliminaries}
In this section, we recall some basic facts 
on the finite length analysis of LDPC codes
under iterative decoding.
We also introduce some notations used throughout this paper.

\subsection{Ensemble and Channel Model}
In this paper, we consider irregular LDPC code ensembles \cite{Irregular}. 
An irregular LDPC code ensemble is defined by the set of
bipartite graphs with variable nodes and check nodes.
Let $\mathcal{L}$ and $\mathcal{R}$ be 
the sets of degrees of variable nodes and check nodes, respectively.
Irregular LDPC code ensembles are characterized with the block length $n$
and two polynomials,
$\lambda(x) = \sum_{i\in\mathcal{L}}\lambda_i x^{i-1}$ and
$\rho(x) = \sum_{i\in\mathcal{R}}\rho_i x^{i-1}$,
where $\lambda_i$ and $\rho_i$ are
the fractions of edges connected to variable nodes and check nodes
of degree $i$, respectively.
The derivatives of $\lambda(x)$ and $\rho(x)$ are
$\lambda'(x) = \sum_{i\in\calL}(i-1)\lambda_i x^{i-2}$
and
$\rho'(x) = \sum_{i\in\calR}(i-1)\rho_i x^{i-2}$,
respectively.

We assume the transmission over the binary erasure channel (BEC)
with channel erasure probability $\epsilon$.

\subsection{Peeling Algorithm}
The peeling algorithm (PA) \cite{Luby97} is 
a sequential iterative decoding algorithm for BEC.
It is know that PA and brief propagation (BP) decoder have
the same decoding result.
A {\it residual graph} at each iteration consists of
variable nodes that are still unknown
and the check nodes and the edges connecting to those variable nodes.
The decoder proceeds as follows.
\paragraph{Initialization}
Variable nodes receive the channel outputs.
The variable nodes receiving the known values
send their values to the check nodes connected to them.
Then the variable nodes sending their values
and edges connecting to those variable nodes
are removed from the graph.
\paragraph{Iteration}
The decoder uniformly chooses a check node of degree one
in the residual graph.
The chosen check node sends 
the value computed from the received values
to the adjacent variable node.
The variable node propagates this value to all adjacent check nodes.
The variable node is removed together with its edges.
\paragraph{Decision}
If the decoder does not find any check nodes of degree one
in the residual graph, then the decoding halts.
If the residual graph is empty,
then the decoding succeeds,
otherwise it fails.

\subsection{Analysis of Residual Graph}
Let $t$ denote the iteration round of PA
and $\xi$ be the total number of edges in the original graph.
We define
\begin{equation}
 \tau 
  :=
 \frac{t}{\xi}. 
 \label{eq:tau}
\end{equation}
Define a parameter $y$ such that 
$\diff{y}/\diff{\tau} = - 1/(\ep \lambda (y))$
and $y = 1$ at $\tau = 0$.
Let $l_{k,t}$ and $r_{i,t}$ denote 
random variables representing the number of edges connecting 
to the variable nodes of degree $k$
and the check nodes of degree $i$, respectively, 
in the residual graph at the iteration round $t$.
%
Let $\rmax$ be the maximum degree of check nodes.
We define 
$\mathcal{\bar{R}} := \{1,2,\dots,\rmax-1\}$.
We also define a set of random variables
\begin{equation}
 \mathcal{D}_{t} 
  :=
 \{l_{k,t} \mid k \in \mathcal{L}\}
 \cup
 \{r_{k,t} \mid k \in \mathcal{\bar{R}}\}. \nonumber
\end{equation}
To simplify the notation, we drop the subscript $t$.
For $X \in \mathcal{D} \cup \{r_{\rmax}\}$, we define $\E{X}(y)$ by
\begin{equation}
 \E{X}(y) := \frac{\mathbb{E}[X]}{\xi}. \nonumber
\end{equation}
For $i \in \calL$ and $j \in \{2,\dots,\rmax\}$
as the block length tends to infinity,
Luby et al.\ \cite{Luby97} showed that 
$\E{X}(y)$ is given by
\begin{IEEEeqnarray}{rCl}
 \E{l}_{i} (y)
  &=&
 \ep \lambda_i y^{i} ,
 \nonumber \\
 \E{r}_{j} (y)
  &=&
 \sum_{i\in\mathcal{R}} \rho_{i} \binom{i-1}{j-1}x^{j} \xbar^{i-j}
 ,
 \nonumber \\
 \E{r}_{1} (y)
   &=&
  x(y-1+\rho(\xbar)),
 \nonumber
\end{IEEEeqnarray}
where $x := \ep \lambda(y)$ and $\xbar:= 1-x$.
We define $\delt{X}{Y}(y)$ by
\begin{equation}
 \delt{X}{Y}(y) := \frac{\mathrm{Cov}[X,Y]}{\xi},
  \hspace{5mm} (X,Y \in\mathcal{D}), \nonumber
\end{equation}
where $\mathrm{Cov}[X,Y]$ is the covariance of $X$ and $Y$.
To simplify the notation, we drop $y$.
In \cite{Amraoui05,Amraoui09},
Amraoui et al.\ showed that
$\delt{X}{Y}$ satisfy 
the following system of differential equations
for irregular LDPC code ensembles
as the block length tends to infinity.
\begin{IEEEeqnarray}{rl}
  \divi{ \delt{X}{Y} }{y}
  =
 -\frac{e}{y}
  \biggl[&
  \sum_{Z \in \mathcal{D}} 
      \Bigl(\rdivi{\hat{f}^{(X)} }{\E{Z}} \delt{Y}{Z}
           +\rdivi{\hat{f}^{(Y)} }{\E{Z}} \delt{X}{Z}\Bigr) 
    \nonumber \\
  &+\hat{f}^{(X,Y)} 
 \biggr], \label{CE}
\end{IEEEeqnarray}
and this system of differential equation 
is referred to as covariance evolution.
Let $I_{\{\cdot\}}$ be the indicator function 
which is 1 if the condition inside the braces is fulfilled 
and 0 otherwise.
Define 
$e(y) := \sum_{i\in\calL}\E{l}_i = xy$, \
$x' := \divi{x}{y}$, \
$a := \sum_{i\in\calL}\frac{i\E{l}_i}{e} =\frac{x'y+x}{x}$
and
$G_{j}(y) := \frac{j(\E{r}_{j+1}-\E{r}_j)}{x}$.
The terms in the covariance evolution are given by the following 
for $k,s \in\calL$ , $i\in\bar{\calR}$ and $j\in\{1,2,\dots,\rmax-2\}$
\begin{IEEEeqnarray}{rCl}
 \rdivi{\hat{f}^{(l_{k})}}{\E{l}_{s}}
   &=& 
 \frac{k\E{l}_k}{e^2} -I_{\{k=s\}}\frac{k}{e}, 
\nonumber \\
  \rdivi{\hat{f}^{(l_{k})}}{\E{r}_i}
  &=& 
  0,  \nonumber \\
 \rdivi{\hat{f}^{(r_j)}}{\E{l}_{k}} 
   &=& 
 -\frac{2a-k-1}{e}\frac{G_j}{y},
 \nonumber \\
 \rdivi{\hat{f}^{(r_j)}}{\E{r}_i}
   &=& 
 j\frac{a-1}{e}(I_{\{i=j+1\}} - I_{\{i=j\}}), 
 \nonumber \\
 \rdivi{\hat{f}^{(r_{\rmax-1})}}{\E{l}_{k}} 
   &=& 
  (\rmax-1)\frac{a-1}{e}
 -\frac{2a-k-1}{e}\frac{G_{\rmax-1}}{y},
  \nonumber \\
 \rdivi{\hat{f}^{(r_{\rmax-1})}}{\E{r}_i}
   &=& 
 -(\rmax-1)\frac{a-1}{e} (1+I_{\{i=\rmax-1\}}),
  \nonumber
\end{IEEEeqnarray}
and for $k,s\in\calL$ and $i,j\in\bar{\calR}$
\begin{IEEEeqnarray}{rl}
 \hat{f}^{(l_{k}, l_{s})}
  &= 
  ks \frac{\E{l}_k}{e}(I_{\{k=s\}} - \frac{\E{l}_s}{e}), 
 \nonumber \\
 \hat{f}^{(l_{k}, r_{i})}
 &= 
 (a-k)\frac{k\E{l}_k}{e}\frac{G_i}{y}, 
  \nonumber \\
 \hat{f}^{(r_{i}, r_{j})}
  &= 
  \frac{x''x-(x')^{2}}{x^{2}}G_i G_j
   \nonumber \\
  +&ij\frac{x'}{x^{2}}
  \bigl[
    I_{\{i=j\}} (\E{r}_{j+1} + \E{r}_{j})
   -I_{\{i=j+1\}} \E{r}_i
   -I_{\{j=i+1\}} \E{r}_j
  \bigr]. 
\nonumber
\end{IEEEeqnarray}

Initial conditions of the covariance evolution are also given by
Amraoui et al.\ \cite{Amraoui05,Amraoui09}.
For $i,j \in \bar{\calR} \cup \{\rmax\}$ and $k,s\in\calL$,
the initial conditions of the covariance evolution are derived as follows:
\begin{IEEEeqnarray}{rCl}
 \delt{l_k}{l_s}(1)
  &=&
 I_{\{k=s\}}k\lambda_k \ep \epbar, 
 \nonumber \\
 \delt{l_k}{r_i}(1)
  &=&
 -k\lambda_k \ep\epbar G_{i}(1),
 \nonumber \\
  \delt{r_i}{r_j}(1) 
  &=&
   I_{\{i=j\}}i \E{r}_{i}(1)
  -V_{i,j}(1)
  +\lambda'(1) \ep\epbar G_{i}(1)G_{j}(1), 
 \nonumber
\end{IEEEeqnarray}
where $\epbar := 1 - \ep$
and
\begin{IEEEeqnarray}{rCl}
 V_{i,j}(y)
  :=
 \sum_{s\in\mathcal{R}}s\rho_s 
   \binom{s-1}{i-1}\binom{s-1}{j-1} x^{i+j}\xbar^{2s-i-j}.
 \nonumber
\end{IEEEeqnarray}

\subsection{Scaling Law}

Let $P_B(\ep,n)$ be the block error probability under BP decoding
for channel erasure probability $\ep$ and block length $n$.
\textit{Threshold} is defined by
\begin{IEEEeqnarray}{C}
 \ep^{*} 
  := 
 \sup \{\ep\in[0,1] \mid \lim_{n\to\infty}P_B(\ep,n) = 0 \},
 \nonumber
\end{IEEEeqnarray}
and characterized via \textit{density evolution} as follows:
\begin{IEEEeqnarray}{C}
 \ep^{*}
  =
 \sup \{\ep\in[0,1] \mid y > 1 - \rho(1-\ep\lambda(y)), \forall y
 \in(0,1]\}.
 \nonumber
\end{IEEEeqnarray}

The curve of the block error probability for finite length LDPC codes
is divided two regions which called
\textit{waterfall region} and \textit{error floor region}.
In the waterfall region,
the block error probability drops off steeply
as the function of channel erasure probability.
In the error floor region,
the block error probability has a gentle slope.
A \textit{scaling law} is a technique
to estimate the waterfall region.
The scaling law is based on the analysis of the residual graphs.

In \cite{Amraoui09}, the block error probability $P_{B}(n,\ep)$ is given by
\begin{IEEEeqnarray}{c}
 P_{B}(n,\ep)
  =
 Q\bigl( \frac{\sqrt{n}(\ep^{*}-\ep)}{\alpha} \bigr)
 + o(1),
 \nonumber
\end{IEEEeqnarray}
where $\alpha$ is {\it slope scaling parameter} depending on the ensemble
and the $Q$-function is defined by
\begin{IEEEeqnarray}{c}
 Q(z) 
  :=
 \frac{1}{\sqrt{2\pi}}
  \int_{z}^{\infty} e^{-\frac{x^2}{2}} \diff{x}.
 \nonumber
\end{IEEEeqnarray}
In \cite{Amraoui09},
the slope scaling parameter is derived as
\begin{equation}
 \alpha 
  =
 -\sqrt{ \frac{n}{\xi} } 
 \sqrt{ \left. \delt{r_1}{r_1} \right|_{\ep^{*};y^{*}} }
 \left( \left.\rdivi{\E{r}_1}{\ep}\right|_{\ep^{*};y^{*}} \right)^{-1}
 \label{eq:alpha}
\end{equation}
where $y^{*}$ is the non-zero solution of $\E{r}_1(y) = 0$ at the threshold
(i.e.\ define $y^{*}$ such that 
$y^{*} = 1- \rho(1-\ep^{*}\lambda(y^{*}))$)
and $\xi$ is the total number of edges in the original graph.
\section{Main Results}
We show, in the following theorem,
the analytical solution of the covariance evolution,
for irregular LDPC code ensembles.
The proof shall be given in Section $\ref{sec_proof}$.

\begin{theorem} \label{theorem1} \upshape
Consider transmission over the BEC($\ep$).
Let $\tau$ be the normalized iteration round of PA
as defined in (\ref{eq:tau}).
A parameter $y$ is defined by 
$dy / d\tau = - 1 /(\ep\lambda(y))$.
For an irregular LDPC code ensemble,
$i,j \in\mathcal{\bar{R}}$ and $k,s \in \mathcal{L}$, 
in the limit of the code length,
we obtain the following.
\begin{IEEEeqnarray}{rCl}
 \delt{l_{k}}{l_{s}}
  &=&
 -\frac{ks \E{l}_{k} \E{l}_{s}}{e^{2}}F
 +\frac{\ep \E{l}_{k} \E{l}_{s}}{e} 
    \bigl[ k(y^{s}-1)  +s(y^{k}-1) \bigr]
  \nonumber \\
 &&+I_{\{k=s\}}
   k \E{l}_{k}(1-\ep y^{k}),
 \label{delt_ll} \\
 \delt{l_s}{r_j}
  &=&
  \bigl[  F \frac{s\E{l}_s}{e} -\ep \E{l}_s (y^{s}-1) \bigr]
   \bigl( \frac{x'}{x}G_j -  I_{\{j=1\}} \bigr)
  \nonumber \\
 &&-\frac{s\E{l}_s}{e}G_j
  (\frac{F' + x}{2} -\ep x y^{s}),
  \label{delt_lr} \\
 \delt{r_i}{r_j}
&=&
-F\bigl(\frac{x'}{x}G_i -I_{\{i=1\}}\bigr)
  \bigl(\frac{x'}{x}G_j -I_{\{j=1\}}\bigr)  
  \nonumber \\
&&+G_i G_j
 \bigl(
    F' \frac{x'}{x} 
  -\sum_{s\in\calL}\ep^{2} s\lambda_{s}y^{2s-2}
  +x^{2} 
 \bigr)
 -V_{i,j}
 \nonumber \\
 &&+\bigl(
   I_{\{j=1\}}G_i
  +I_{\{i=1\}} G_j
     \bigr)
  \bigl[
   x (e-x)
  -\frac{F'-x}{2}
  \bigr]
  \nonumber \\
 &&+I_{\{i=j\}}i \E{r}_{i}
+I_{\{i=j=1\}}(e-x)^{2},
 \label{delt_rr}
\end{IEEEeqnarray}
where
$
 F 
:=
 \sum_{i}\frac{\lambda_{i}}{i}
  [\ep^{2}(y^{i}-1)^{2} + \ep(y^{i}-1)]
$ 
and
$
 F'
  =
 \divi{F}{y}
  =
 2\sum_{i}\ep^{2}\lambda_{i}y^{2i-1}
 -(\ep-\epbar)x
$.
\end{theorem}

Using Theorem \ref{theorem1}, we can obtain the following corollary.
\begin{corollary} \label{col1} \upshape
 Let $\ep^{*}$ be the threshold of the ensemble under BP decoding,
 $n$ be the block length
 and $\xi$ be the total number of edges in the original graph.
 For irregular LDPC codes, the slope scaling parameter $\alpha$ is given by
\begin{IEEEeqnarray}{rCl}
 \alpha 
  &=&
  \Bigl[
    \frac{ \rho(\xbar^{*})^{2}
    -\rho(\xbar^{*2})
    -\xbar^{*2} \rho'(\xbar^{*2})
         }
         {\rho'(\xbar^{*})^{2}}
    +\frac{1
        -2x^{*}\rho(\xbar^{*})}{\rho'(\xbar^{*})}
     \nonumber \\
   &&{}+x^{*2}
        -\ep^{*2}\lambda(y^{*2})
        -\ep^{*2} y^{*2} \lambda'(y^{*2})
  \Bigr]^{\frac{1}{2}}
   \sqrt{\frac{n}{\xi}}
    \frac{1}{\lambda(y^{*})},
   \label{eq:col1}
\end{IEEEeqnarray}
where
$x^{*} := \ep^{*}\lambda(y^{*})$
and $\xbar^{*} := 1 - x^{*}$.
\end{corollary}
\begin{IEEEproof}
Since 
$\left.\E{r}_1\right|_{\ep^{*};y^{*}} = 0$
and 
$\left.\rdivi{\E{r}_1}{y}\right|_{\ep^{*};y^{*}} = 0$,
we see that
$1-y^{*} = \rho(\xbar^{*})$
and
$\rho'(\xbar^{*})\ep^{*}\lambda'(y^{*}) = 1$.
Using those equations,
we have from (\ref{delt_rr}),
\begin{IEEEeqnarray}{rl}
 \left.\delt{r_1}{r_1} \right|&_{\ep^{*};y^{*}}
  \nonumber \\
  =& 
  x^{*2}
  [\rho(\xbar^{*})^{2} - \xbar^{*2}\rho'(\xbar^{*2}) - \rho(\xbar^{*2})]
   \nonumber \\
 &+x^{*2}\rho'(\xbar^{*})
  [1-2x^{*}\rho(\xbar^{*})] \nonumber \\
 &+(x^{*}\rho'(\xbar^{*}))^{2}
  [x^{*2} - \ep^{*2}\lambda'(y^{*2})y^{*2} - \ep^{*2}\lambda(y^{2*})].
 \nonumber
\end{IEEEeqnarray}
Recall that 
$\E{r}_1 = x(y-1+\rho(\xbar))$.
We see that
\begin{eqnarray}
 \left.\rdivi{\E{r}_1}{\ep}\right|_{\ep^{*};y^{*}}
  =
 -\lambda(y^{*}) x^{*} \rho'(\xbar^{*}).
 \nonumber
\end{eqnarray}
From (\ref{eq:alpha}), we can obtain (\ref{eq:col1}).
\end{IEEEproof}

\begin{remark} \upshape
 The result of Corollary \ref{col1} is the same as the result 
in \cite{Amraoui07} for irregular LDPC code ensembles.
In particular, for $(d_{v},d_{c})$-regular LDPC code ensembles 
we can write
\begin{eqnarray}
 \alpha
  =
 \ep^{*}\sqrt{\frac{d_{v}-1}{d_{v}}(\frac{1}{x^{*}}-\frac{1}{y^{*}})}.
 \nonumber
\end{eqnarray}
\end{remark}
\section{Lemmas and Proofs} \label{sec_proof}
In this section, we state three lemmas 
and prove Theorem \ref{theorem1}.
Section \ref{proof_delt_ll}, \ref{proof_delt_lr} and \ref{proof_delt_rr}
give (\ref{delt_ll}), (\ref{delt_lr}) and (\ref{delt_rr}),
respectively.

\subsection{Lemma and Proof of (\ref{delt_ll})}\label{proof_delt_ll}
In this section, we give a lemma to prove (\ref{delt_ll})
and we prove (\ref{delt_ll}).

\subsubsection{Lemma to Prove (\ref{delt_ll})}
\begin{lemma} \label{lemma1} \upshape
 Define 
 $U^{(l_k;l_s)}
   :=
  \frac{\delt{l_k}{l_k}}{(k\E{l}_k)^{2}}
 -\frac{\delt{l_s}{l_s}}{(s\E{l}_s)^{2}}$.
For $k,s \in\calL$, we have the following equations.
 \begin{IEEEeqnarray}{l}
  \sum_{k,s\in\calL}\frac{\delt{l_k}{l_s}}{ks}
   =
  \ep\epbar \sum_{i\in\calL}\frac{\lambda_{i}}{i},
  \label{lem1.1}
   \\
  2\frac{\delt{l_k}{l_s}}{k s \E{l}_k \E{l}_s}
 -\frac{\delt{l_k}{l_k}}{(k\E{l}_k)^2}
 -\frac{\delt{l_s}{l_s}}{(s\E{l}_s)^2}
   \nonumber \\
  \hspace{15mm} =
  \bigl(
     \frac{\ep y^{k}-1}{k\E{l}_k}
    +\frac{\ep y^{s}-1}{s\E{l}_s}
  \bigr)
  I_{\{k\neq s\}},
  \label{lem1.2}
   \\
  U^{(l_k;l_s)}
   =
 -\frac{\ep y^{k} -1}{k \E{l}_k}
 +\frac{\ep y^{s} -1}{s \E{l}_s}
  \nonumber \\
 \hspace{15mm}+\frac{2\ep}{e}
  \bigl(
    \frac{y^{k}-1}{k}
   -\frac{y^{s}-1}{s}
  \bigr).
  \label{lem1.3}
 \end{IEEEeqnarray}
\end{lemma}
\begin{IEEEproof}
Define $\delt{l_k}{l_{\Sigma}} = \sum_{s\in\calL}\delt{l_k}{l_s}$.
From the covariance evolution, we have
\begin{IEEEeqnarray}{rCl}
 \divi{\delt{l_k}{l_s}}{y}
  &=&
 -x
  \Bigl(
    \frac{s\E{l}_s}{e^{2}} \delt{l_k}{l_{\Sigma}}
   +\frac{k\E{l}_k}{e^{2}} \delt{l_s}{l_{\Sigma}}
   -\frac{k+s}{e}      \delt{l_k}{l_s}
 \Bigr)
  \nonumber \\
 &&{}-x    \hat{f}^{(l_k,l_s)}.
  \label{CE_ll}
\end{IEEEeqnarray}

\paragraph{Proof of (\ref{lem1.1})}
From (\ref{CE_ll}), we have the following equation:
\begin{eqnarray}
 \sum_{k,s\in\calL} \frac{1}{ks}\divi{\delt{l_k}{l_s}}{y}
  =
 0.
 \nonumber
\end{eqnarray}
From initial conditions, we have
\begin{eqnarray}
 \sum_{k,s\in\calL} \frac{1}{ks} \delt{l_k}{l_s}
  =
 \ep\epbar \sum_{i\in\calL}\frac{\lambda_{i}}{i}.
 \nonumber
\end{eqnarray}
This leads to (\ref{lem1.1}).

\paragraph{Proof of (\ref{lem1.2})}
Obviously we can get (\ref{lem1.2}) for $k=s$.
From (\ref{CE_ll}), we have
\begin{IEEEeqnarray}{rCl}
 \divi{}{y}\bigl( \frac{\delt{l_k}{l_s}}{ks \E{l}_k \E{l}_s} \bigr)
  &=&
 \frac{1}{ks \E{l}_k \E{l}_s}     \divi{\delt{l_k}{l_s}}{y}
-\frac{k+s}{ks \E{l}_k \E{l}_s y} \delt{l_k}{l_s}
 \nonumber \\
  &=&
 -x
 \Bigl(
    \frac{ \hat{f}^{(l_k,l_s)}    }{ks \E{l}_k \E{l}_s} 
   +\frac{ \delt{l_k}{l_{\Sigma}} }{k \E{l}_k e^{2}} 
   +\frac{ \delt{l_s}{l_{\Sigma}} }{s \E{l}_s e^{2}} 
 \Bigr).
 \label{eq:pr1.2.1}
\end{IEEEeqnarray}
From those equations, we have
\begin{IEEEeqnarray}{l}
 \divi{}{y}
  \Bigl(
    2\frac{ \delt{l_k}{l_s} }{ks \E{l}_k \E{l}_s} 
   - \frac{ \delt{l_k}{l_k} }{(k \E{l}_k)^{2}} 
   - \frac{ \delt{l_s}{l_s} }{(s \E{l}_s)^{2}} 
  \Bigr)
   \nonumber \\
  \hspace{15mm}=
 -2\frac{x \hat{f}^{(l_k,l_s)}}{ks \E{l}_k \E{l}_s}
 + \frac{x \hat{f}^{(l_k,l_k)}}{k^{2} \E{l}_k^{2}}
 + \frac{x \hat{f}^{(l_s,l_s)}}{s^{2} \E{l}_s^{2}}
 \nonumber \\
  \hspace{15mm}=
 \frac{1}{y} \bigl( \frac{1}{\E{l}_k} + \frac{1}{\E{l}_s} \bigr),
 \nonumber
\end{IEEEeqnarray}
for $j\neq k$.
This differential equation can be solve as follow:
\begin{IEEEeqnarray}{l}
  2\frac{ \delt{l_k}{l_s} }{ks \E{l}_k \E{l}_s} 
 - \frac{ \delt{l_k}{l_k} }{(k \E{l}_k)^{2}} 
 - \frac{ \delt{l_s}{l_s} }{(s \E{l}_s)^{2}} 
  =
 -\frac{1}{k \E{l}_k}
 -\frac{1}{s \E{l}_s}
 +C,
\nonumber
\end{IEEEeqnarray}
with a constant $C$ which can be determined from initial conditions.
From initial conditions, we get
\begin{align}
 C
 =
  \frac{1}{k\lambda_{k}}
 +\frac{1}{s\lambda_{s}}.
 \nonumber
\end{align}
Thus we have for $k\neq s$
\begin{align}
  2\frac{ \delt{l_k}{l_s} }{ks \E{l}_k \E{l}_s} 
 - \frac{ \delt{l_k}{l_k} }{(k \E{l}_k)^{2}} 
 - \frac{ \delt{l_s}{l_s} }{(s \E{l}_s)^{2}} 
  =&
  \frac{\ep y^{k} - 1}{k\E{l}_k}
 +\frac{\ep y^{s} - 1}{s\E{l}_s}.
 \nonumber
\end{align}
This leads to (\ref{lem1.2}).

\paragraph{Proof of (\ref{lem1.3})}
From (\ref{lem1.2}), we have for all $k,s\in\calL$
\begin{IEEEeqnarray}{rCl}
 \delt{l_k}{l_s}
  &=&
  \bigl[
    \frac{s \E{l}_s}{2} (\ep y^{k} - 1) 
   +\frac{k \E{l}_k}{2} (\ep y^{s} - 1) 
  \bigr]
  I_{\{k \neq s\}}
  \nonumber \\
 &&{}+\frac{s \E{l}_s}{2 k \E{l}_k} \delt{l_k}{l_k}
 +\frac{k \E{l}_k}{2 s \E{l}_s} \delt{l_s}{l_s}.
 \nonumber 
\end{IEEEeqnarray}
The sum of this equation for $s\in\calL$ is written as follows
\begin{IEEEeqnarray}{rCl}
 \delt{l_k}{l_{\Sigma}}
  &=&
   \frac{ae}{2} (\ep y^{k} - 1)
 + \frac{k\E{l}_k}{2} \sum_{s\in\calL} (\ep y^{s} - 1)
 -k \E{l}_k (\ep y^{k} - 1)
  \nonumber \\
 &&{}+ \frac{ae}{2 k\E{l}_k} \delt{l_k}{l_k}
 +k\E{l}_k \sum_{s\in\calL} \frac{ \delt{l_s}{l_s} }{2 s\E{l}_s}.
 \nonumber
\end{IEEEeqnarray}
Combining (\ref{eq:pr1.2.1}) with this equation, we have
\begin{IEEEeqnarray}{rCl}
 \divi{}{y}\Bigl(\frac{\delt{l_k}{l_k}}{(k \E{l}_k)^{2}}\Bigr)
  &=&
  K^{(l_k,l_k)}
 -\frac{x}{e^{2}}  \sum_{s\in\calL}  (\ep y^{s} - 1)
  -\frac{x}{e^{2}}  \sum_{s\in\calL} \frac{\delt{l_s}{l_s}}{s\E{l}_s} 
   \nonumber \\
 &&{}-\frac{a}{y}   \frac{ \delt{l_k}{l_k} }{(k\E{l}_k)^{2}} ,
\nonumber
\end{IEEEeqnarray}
where
\begin{align}
 K^{(l_k,l_k)}
  :=
 -x\frac{\hat{f}^{(l_k,l_k)}}{(k\E{l}_k)^{2}} 
 -\frac{a}{k\E{l}_k y}  (\ep y^{k} - 1)
 +\frac{2}{e y}     (\ep y^{k} - 1).
 \nonumber
\end{align}
From this equation, we have
\begin{align}
 \divi{U^{(l_k;l_s)}}{y}
  =&
   K^{(l_k,l_k)}
  -K^{(l_s,l_s)}
  -\frac{a}{y} U^{(l_k;l_s)}.
 \label{eq_proof1.1}
\end{align}
Note that
\begin{align}
 \int \frac{a}{y} \diff{y}
   =
 \log xy.
\nonumber
\end{align}
Since (\ref{eq_proof1.1}) is a first order differential equation,
it can be solved as follows:
\begin{align}
 U^{(l_k;l_s)}
  =&
 \frac{1}{e}
 \int e
   \bigl( K^{(l_k,l_k)} - K^{(l_s,l_s)} \bigr) \diff{y}
 +\frac{1}{e}C,
\nonumber
\end{align}
with a constant $C$ which is determined from initial conditions.
Note that
\begin{IEEEeqnarray}{l}
 \int e K^{(l_k,l_k)} \diff{y}
   \nonumber \\
 \hspace{5mm}=
 \int 
  \Bigl[
   -\frac{x'y+x}{k\lambda_{k}}
   +\frac{x'y - (k-1)x}{k\E{l}_k}
   +2\ep y^{k-1}
   -\frac{1}{y}
  \Bigr]
  \diff{y}
 \nonumber \\
  \hspace{5mm}=
-\frac{e}{k\lambda_k}
+\sum_{i\in\calL} \frac{\E{l}_i}{k\E{l}_k}I_{\{i\neq k\}}
+\frac{2\ep}{k}y^{k}
-\log y.
 \nonumber
\end{IEEEeqnarray}
We get
\begin{IEEEeqnarray}{rl}
 U^{(l_k;l_s)}
  =&
  -\frac{\ep y^k -1}{k \E{l}_{k}} 
  +\frac{\ep y^s -1}{s \E{l}_{s}}
   \nonumber \\
 &{}+\frac{1}{e}
  \Bigl(
    \frac{2\ep y^{k} -1}{k}
   -\frac{2\ep y^{s} -1}{s}
  + C
  \Bigr).
\nonumber
\end{IEEEeqnarray}
From initial conditions, we have
$ U^{(l_k;l_s)}(1)
  = 
 \frac{\epbar}{\ep}
 \bigl(
   \frac{1}{k\lambda_{k}}
  -\frac{1}{s\lambda_{s}}
 \bigr)$
and
$ C 
  = 
  \frac{1-2\ep}{k}
 -\frac{1-2\ep}{s}$.
Therefore we have
\begin{IEEEeqnarray}{l}
 U^{(l_k;l_s)}
   =
  \frac{1 - \ep y^{k}}{k \E{l}_{k}} 
 -\frac{1 - \ep y^{s}}{s \E{l}_{s}}
 +\frac{2\ep}{e}
  \Bigl(
    \frac{y^{k} - 1}{k}
   -\frac{y^{s} - 1}{s}
  \Bigr).
\nonumber
\end{IEEEeqnarray}
This leads (\ref{lem1.3}).
\end{IEEEproof}

\subsubsection{Proof of (\ref{delt_ll})}
By definition of $U^{(l_k;l_s)}$, we have
\begin{align}
 \E{l}_k \delt{l_s}{l_s}
  =
 (s\E{l}_s)^{2} 
  \bigl(
    \frac{\delt{l_k}{l_k}}{k^{2}\E{l}_k}
   -\E{l}_k U^{(l_k;l_s)}
  \bigr).
 \nonumber
\end{align}
The sum of this equation for $k\in\calL$ is written as follows:
\begin{eqnarray}
  e \delt{l_s}{l_s}
   =
  (s\E{l}_s)^{2}
  \sum_{k\in\calL}
  \bigl(
    \frac{\delt{l_k}{l_k}}{k^{2}\E{l}_k}
   -\E{l}_k U^{(l_k;l_s)}
  \bigr).
 \label{lem1.7}
\end{eqnarray}
From (\ref{lem1.2}), we see that for all $k,s\in\calL$
 \begin{IEEEeqnarray}{l}
  \frac{1}{2} \frac{\E{l}_s}{k^{2} \E{l}_k} \delt{l_k}{l_k}
 +\frac{1}{2} \frac{\E{l}_k}{s^{2} \E{l}_s} \delt{l_s}{l_s}
   \nonumber \\
  \hspace{5mm}=
  \frac{1}{ks} \delt{l_k}{l_s}
 -\frac{1}{2} \E{l}_s \frac{\ep y^{k} - 1}{k}I_{\{k \neq s\}}
 -\frac{1}{2} \E{l}_k \frac{\ep y^{s} - 1}{s}I_{\{k \neq s\}}.
  \nonumber
 \end{IEEEeqnarray}
The sum over this equation for $k,s\in\calL$ 
is written as follows:
 \begin{eqnarray}
  e\sum_{k\in\calL} \frac{\delt{l_k}{l_k}}{k^{2} \E{l}_k} 
  =
  \sum_{k,s\in\calL} \frac{\delt{l_k}{l_s}}{ks} 
 +\sum_{k\in\calL} (\E{l}_k -e) \frac{\ep y^{k} - 1}{k}.
   \label{lem1.5}
\end{eqnarray}
Combining (\ref{lem1.5}) with (\ref{lem1.1}), we have
\begin{eqnarray}
  \sum_{k\in\calL} \frac{\delt{l_k}{l_k}}{k^{2} \E{l}_k} 
  =
  \frac{\ep\epbar}{e} \sum_{k\in\calL} \frac{\lambda_{k}}{k}
 +\sum_{k\in\calL} \frac{\E{l}_k -e}{e} \frac{\ep y^{k} - 1}{k}.
 \label{lem1.6}
\end{eqnarray}
From (\ref{lem1.3}), we have
\begin{IEEEeqnarray}{rCl}
 \sum_{k\in\calL}\E{l}_{k} U^{(l_k,l_s)}
  &=&
  \frac{e}{s \E{l}_s}(\ep y^{s} -1)
 - 2\ep \frac{y^{s}-1}{s}
   \nonumber \\
 &&{}-\sum_{k\in\calL} \frac{\ep y^{k} -1}{k}
 +\frac{2\ep}{e}
  \sum_{k\in\calL}\frac{\E{l}_k (y^{k}-1)}{k}.
 \label{lem1.8}
\end{IEEEeqnarray}
Combining (\ref{lem1.7}) with (\ref{lem1.6}) and (\ref{lem1.8}),
we obtain
\begin{eqnarray}
 \delt{l_{s}}{l_{s}}
  =
 -\frac{(s \E{l}_{s})^{2}}{e^{2}}F
 +2\ep \frac{s \E{l}_{s}^{2} }{e} (y^{s}-1)  
 +s \E{l}_{s}(1-\ep y^{s}).
 \nonumber
\end{eqnarray}
From this equation and (\ref{lem1.2}),
we can obtain (\ref{delt_ll}) for $k,s\in\calL$.

\subsection{Lemma and Proof of (\ref{delt_lr})}\label{proof_delt_lr}
In this section, we introduce a lemma to prove (\ref{delt_lr})
and we prove (\ref{delt_lr}).
\subsubsection{Lemma to Prove (\ref{delt_lr})}
\begin{lemma} \label{lemma2} \upshape
Define 
$A^{(l_{\Sigma},r_{j})} := \sum_{i\in\calL}\frac{1}{i}\delt{l_i}{r_j}$,
$A^{(l_{\Sigma},r_{\Sigma})} := \sum_{j\in\bar{\calR}}A^{(l_{\Sigma},r_{j})}$,
$S^{(l_i,l_s;r_j)} 
  := 
 \frac{1}{i\E{l}_i}\delt{l_i}{r_j}
-\frac{1}{s\E{l}_s}\delt{l_s}{r_j}$,
$S^{(l_i,l_s;r_\Sigma)} := \sum_{j\in\bar{\calR}}S^{(l_i,l_s;r_j)} $
and
$G_{\Sigma} 
 := \sum_{j\in\bar{\calR}}G_{j}
 = \frac{\rmax\E{r}_{\rmax} -e}{x}$.
For $j\in\bar{\calR}$ and $k,s\in\calL$, we have the following equations.
 \begin{IEEEeqnarray}{L}
  A^{(l_{\Sigma},r_{\Sigma})}
   =
  \ep \epbar \sum_{i\in\calL}\frac{\lambda_i}{i}(y^{i}-1)
   (G_{\Sigma}\frac{x'}{x}-1)
 -\epbar x G_{\Sigma},
  \label{lem2.1} \\
  A^{(l_{\Sigma},r_{j})}
   =
  \ep \epbar \sum_{i\in\calL}\frac{\lambda_i}{i}(y^{i}-1)
   (G_j\frac{x'}{x}-I_{\{j=1\}})
 -\epbar x G_j,
  \label{lem2.2} \\
  S^{(l_k,l_s;r_{\Sigma})}
   = 
  -\ep\bigl(G_{\Sigma} \frac{x'}{x} - 1 \bigr)
      \bigl(  \frac{y^{k}-1}{k} - \frac{y^{s}-1}{s}  \bigr)
   \nonumber \\
  \hspace{18mm}{}+\ep G_{\Sigma}   (y^{k-1}-y^{s-1}),
  \label{lem2.3} \\
  S^{(l_k,l_s;r_{j})}
   =
  -\ep \bigl(G_j \frac{x'}{x} - I_{\{j=1\}} \bigr)
       \bigl(  \frac{y^{k}-1}{k} - \frac{y^{s}-1}{s}  \bigr)
   \nonumber \\
  \hspace{18mm}{}+\ep G_j    (y^{k-1}-y^{s-1}).
   \label{lem2.4}
 \end{IEEEeqnarray}
\end{lemma}
We use (\ref{lem2.1}) and (\ref{lem2.3}) to prove the basis of
the mathematical induction in proof of (\ref{lem2.2}) and
(\ref{lem2.4}),
respectively.
\begin{IEEEproof}
First, we will derive differential equations.
We define
$\delt{l_{\Sigma}}{r_j}
 :=
\sum_{k\in\mathcal{L}}\delt{l_k}{r_j}$,
$\delt{l_{k}}{r_{\Sigma}}
 :=
\sum_{j\in\mathcal{\bar{R}}} \delt{l_{k}}{r_{j}}$
and 
$\delt{l_k}{r_{\rmax}} 
 :=
 \delt{l_{k}}{l_{\Sigma}} - \delt{l_{k}}{r_{\Sigma}}$,
respectively.
From the covariance evolution (\ref{CE}),
we can write for $j \in \bar{\calR}$ and $k\in \calL$ 
\begin{IEEEeqnarray}{RL}
 \divi{\delt{l_k}{r_j}}{y}
 =&
 D^{(l_k,r_j)}
  -\frac{k\E{l}_k}{ey} \delt{l_{\Sigma}}{r_j}
  +\frac{k}{y} \delt{l_k}{r_j}
   \nonumber \\
 &{}-j\frac{x'}{x}(\delt{l_k}{r_{j+1}}   -\delt{l_k}{r_j})
   ,
\label{diff_lr}
\end{IEEEeqnarray}
where 
\begin{IEEEeqnarray}{RCL}
 D^{(l_k,r_j)}
  &:=&
 2\frac{x'}{e}G_{j} \delt{l_k}{l_{\Sigma}}
   -\frac{G_{j}}{y^{2}}
    \sum_{i\in\mathcal{L}} (i-1)\delt{l_{k}}{l_{i}}
  \nonumber \\
  &&-x\hat{f}^{(l_{k},r_{j})}.
 \nonumber
\end{IEEEeqnarray}
We define 
$A^{(l_{\Sigma}, r_j)} 
 := 
 \sum_{k\in\mathcal{L}} \frac{1}{k}\delt{l_k}{r_j}$,~
$A^{(l_{\Sigma},r_{\Sigma})}
 :=
\sum_{j\in\mathcal{\bar{R}}} A^{(l_{\Sigma},r_{j})}$
and
$D^{(l_k,r_{\Sigma})} := \sum_{j\in\bar{\calR}} D^{(l_k,r_j)}
$.
From  (\ref{diff_lr}), we have for $k\in\bar{\calR}$
\begin{IEEEeqnarray}{L}
 \divi{ A^{(l_{\Sigma}, r_j)} }{y}
  =
  \sum_{k\in\calL}\frac{D^{(l_k,r_j)}}{k}
 - j\frac{x'}{x}
     \bigl(
       A^{(l_{\Sigma},r_{j+1})}
      -A^{(l_{\Sigma},r_{j})}
     \bigr).
\label{diff_A}
\end{IEEEeqnarray}
The sum over this equation for $j\in\bar{\calR}$ 
is written as the follows:
\begin{IEEEeqnarray}{RL}
 \divi{A^{(l_{\Sigma},r_{\Sigma})}}{y}
  =&
   \sum_{k\in\calL}\frac{D^{(l_k,r_{\Sigma})}}{k}
  -(\rmax-1)\frac{x'}{x}
      \sum_{k\in\mathcal{L}}\frac{1}{k}\delt{l_k}{l_{\Sigma}}
   \nonumber \\
  &{}+\rmax\frac{x'}{x}     A^{(l_{\Sigma},r_{\Sigma})}.
 \label{diff_As}
\end{IEEEeqnarray}

From (\ref{diff_lr}), we see that
\begin{IEEEeqnarray}{rCl}
 \divi{}{y}\Bigl(\frac{\delt{l_k}{r_j}}{k\E{l}_k}\Bigr)
  &=&
    \frac{D^{(l_k,r_j)}}{k\E{l}_k}
   -\frac{1}{ey}\delt{l_{\Sigma}}{r_j}
  \nonumber \\
   &&{}-j\frac{x'}{x}\frac{\delt{l_k}{r_{j+1}} -\delt{l_k}{r_j}}{k\E{l}_k}.
 \label{diff_lr_sub}
\end{IEEEeqnarray}
Define 
$S^{(l_k,l_s;r_j)} 
 := 
  \frac{\delt{l_k}{r_j}}{k\E{l}_k}
 -\frac{\delt{l_s}{r_j}}{s\E{l}_s}$,
$S^{(l_k,l_s;l_{i})}
 :=
 \frac{\delt{l_k}{l_{i}}}{k\E{l}_k}
-\frac{\delt{l_s}{l_{i}}}{s\E{l}_s}$
and
$S^{(l_k,l_s;l_{\Sigma})}
 :=
 \frac{\delt{l_k}{l_{\Sigma}}}{k\E{l}_k}
-\frac{\delt{l_s}{l_{\Sigma}}}{s\E{l}_s}$.
From (\ref{diff_lr_sub}), we have
\begin{IEEEeqnarray}{RCL}
 \divi{S^{(l_k,l_s;r_j)}}{y}
  &=&
  \frac{D^{(l_k,r_j)}}{k\E{l}_k}
  -\frac{D^{(l_s,r_j)}}{s\E{l}_s}
  \nonumber \\
 &&{}-j\frac{x'}{x}
     \bigl(
         S^{(l_k,l_s;r_{j+1})}
        -S^{(l_k,l_s;r_{j})}           
     \bigr),
\label{diff_lrS}
\end{IEEEeqnarray}
for $k,s \in \calL$ and $j\in\bar{\calR}$.
The sum over this equation for $j\in\bar{\calR}$
is written as the follows:
\begin{IEEEeqnarray}{RCL}
 \divi{S^{(l_k,l_s;r_{\Sigma})}}{y}
  &=&
   \frac{D^{(l_k,r_{\Sigma})}}{k\E{l}_k}
  -\frac{D^{(l_s,r_{\Sigma})}}{s\E{l}_s}
    -(\rmax-1)\frac{x'}{x} S^{(l_k,l_s;l_{\Sigma})}  
   \nonumber \\
  &&+\rmax    \frac{x'}{x} S^{(l_k,l_s;r_{\Sigma})}
 .
\label{diff_lrSs}
\end{IEEEeqnarray}

\paragraph{Proof of (\ref{lem2.1})}
Since (\ref{diff_As}) is a first order differential equation, 
it can be solve as follows
\footnote{In a way similar to Section \ref{sec_der_lem2.2},
we perform this calculation.}:
\begin{IEEEeqnarray}{L}
 A^{(l_{\Sigma},r_{\Sigma})}
 \nonumber \\
 =
 x^{\rmax} \int \frac{1}{x^{\rmax}} 
  \bigl[
  \sum_{k\in\calL}\frac{D^{(l_k,r_{\Sigma})}}{k}
    -(\rmax-1)\frac{x'}{x}\sum_{k\in\calL}\frac{\delt{l_k}{l_{\Sigma}}}{k}
  \bigr] 
 \diff{y}
  \nonumber \\
\hspace{5mm}{}+C x^{\rmax}
 \nonumber \\
 = 
 \ep\epbar \sum_{k\in\calL}\frac{\lambda_{k}}{k}(y^{k}-1) 
 \bigl(  G_{\Sigma}\frac{x'}{x} - 1 \bigr)
  + \epbar xy
 + C x^{\rmax},
 \nonumber
\end{IEEEeqnarray}
with a constant $C$ which is determined initial conditions.
From initial conditions, we see that
\begin{align}
  A^{(l_{\Sigma},r_{\Sigma})}(1)
  =&
 \ep\epbar(1-\rmax\rho_{\rmax}\ep^{\rmax-1}).
\nonumber
\end{align}
From this equation, we can determine 
$ C  = -\rmax\rho_{\rmax}\epbar $.
Thus, we get
\begin{IEEEeqnarray}{L}
 A^{(l_{\Sigma},r_{\Sigma})}
=
 \ep\epbar
  \sum_{k\in\calL}\frac{\lambda_{k}}{k}(y^{k}-1) 
 \bigl(  G_{\Sigma}\frac{x'}{x} - 1 \bigr)
 -\epbar x G_{\Sigma}.
\nonumber
\end{IEEEeqnarray}
Hence, we have (\ref{lem2.1}).

\paragraph{Proof of (\ref{lem2.2})} \label{sec_der_lem2.2}
Since (\ref{diff_A}) is a first order differential equation,
it can be solve as follows:
\begin{IEEEeqnarray}{RCL}
 A^{(l_{\Sigma},r_{j})}
  &=&
 x^{j} \int \frac{1}{x^{j}} 
  \bigl(
    \sum_{k\in\calL}\frac{D^{(l_k,r_j)}}{k}
   -j\frac{x'}{x}A^{(l_{\Sigma},r_{j+1})}
  \bigr) 
 \diff{y}
  \nonumber \\
 &&{}+ C_{l_{\Sigma},r_{j}} x^{j},
 \label{int_A}
\end{IEEEeqnarray}
with a constant $C_{l_{\Sigma},r_{j}}$ 
which can be determined from initial conditions.

We solve (\ref{int_A}) by mathematical induction
for $j \in\{2,3,\dots,\rmax-1\}$.
From (\ref{lem2.1}), we have
\begin{IEEEeqnarray}{rCl}
 A^{(l_{\Sigma},r_{\rmax})}
  &=&
  \ep\epbar \sum_{i\in\calL} \frac{\lambda_i}{i}(y^i -1)\frac{x'}{x}
  G_{\rmax}
  -\epbar xG_{\rmax}
  \nonumber
\end{IEEEeqnarray}
where $G_{\rmax} = -\frac{\rmax\E{r}_{\rmax}}{x}$.
Using the same method in the induction step,
we can show that $A^{(l_{\Sigma},r_{\rmax-1})}$ fulfill
(\ref{lem2.2}).

We show that if $A^{(l_{\Sigma},r_{j+1})}$ fulfill (\ref{lem2.2}),
then also $A^{(l_{\Sigma},r_{j})}$ fulfill (\ref{lem2.2}).
Using the induction hypothesis, we have
\begin{IEEEeqnarray}{L}
 \sum_{k\in\calL}\frac{D^{(l_k,r_j)}}{k}
 -j\frac{x'}{x}A^{(l_{\Sigma},r_{j+1})}
  \nonumber \\
 \hspace{5mm}=
  -j\frac{(x')^{2}}{x^{2}}
    \ep \epbar \sum_{i\in\calL}\frac{\lambda_{i}}{i}(y^{i}-1)
     G_{j+1}
  +j x'  G_{j+1} \epbar
  \nonumber  \\
 \hspace{9mm}{}+x' G_j   \epbar 
  +G_j     \ep\epbar
     \sum_{i\in\mathcal{L}}\frac{\lambda_{i}}{i} (y^{i}-1)
      (\frac{x''}{x}-2\frac{(x')^{2}}{x^{2}}).
 \nonumber
\end{IEEEeqnarray}
Using integration by parts, we have
\begin{IEEEeqnarray}{L}
\int \frac{1}{x^{j}}
  G_j  \ep\epbar
     \sum_{i\in\mathcal{L}}\frac{\lambda_{i}}{i} (y^{i}-1)
      \frac{x''}{x} \diff{y}
 \nonumber \\
 \hspace{5mm}=
 \ep\epbar\frac{G_{j}}{x^{j+1}}
     \sum_{i\in\mathcal{L}}\frac{\lambda_{i}}{i} (y^{i}-1)
      x' 
  \nonumber \\
 \hspace{9mm}-\ep\epbar
  \int\Bigl(\frac{G_{j}}{x^{j+1}}
     \sum_{i\in\mathcal{L}}\frac{\lambda_{i}}{i} (y^{i}-1)\Bigr)'
      x'  \diff{y}.
 \label{lr_a_1}
\end{IEEEeqnarray}
Note that
$ G_j'
  =
 -j\frac{x'}{x}G_{j+1}
 +(j-1)\frac{x'}{x}G_{j} $
for $j\in\{2,\dots,\rmax-1\}$.
From (\ref{lr_a_1}), we have
\begin{IEEEeqnarray}{L}
\int 
  \frac{1}{x^{j}}
     \Bigl[
   -\ep\epbar j\frac{(x')^{2}}{x^{2}} G_{j+1}
       \sum_{i\in\mathcal{L}}\frac{\lambda_{i}}{i} (y^{i}-1)
   +\sum_{i\in\calL} \frac{D^{(l_i,r_j)}}{i}
     \Bigr]\diff{y}
 \nonumber \\
 \hspace{5mm}=
 \ep\epbar G_{j}
     \sum_{i\in\mathcal{L}}\frac{\lambda_{i}}{i} (y^{i}-1)
      \frac{x'}{x^{j+1}}.
 \label{cal4}
\end{IEEEeqnarray}
We get
\begin{align}
 \int \frac{1}{x^{j}}
  jG_{j+1}x' \epbar \diff{y}
 =
 -\frac{\epbar}{x^{j-1}}G_j.
 \label{cal5}
\end{align}
From the sum over (\ref{cal4}) and (\ref{cal5}), we have
\begin{IEEEeqnarray}{L}
 x^{j} \int \frac{1}{x^{j}} 
  \bigl(
    \sum_{i\in\calL}\frac{D^{(l_i,r_j)}}{i}
   -j\frac{x'}{x}A^{(l_{\Sigma},r_{j+1})}
  \bigr)
 \diff{y}
  \nonumber \\
  \hspace{5mm}=
 \ep\epbar 
   \sum_{i\in\mathcal{L}}\frac{\lambda_{i}}{i} (y^{i}-1)
      G_{j} \frac{x'}{x}
 -\epbar G_j x.
\nonumber
\end{IEEEeqnarray}
Thus, we have
\begin{align}
 A^{(l_{\Sigma},r_{j})}
  =
 \ep\epbar 
   \sum_{i\in\mathcal{L}}\frac{\lambda_{i}}{i} (y^{i}-1)
      G_{j} \frac{x'}{x}
 -\epbar G_j x
 +C_{l_{\Sigma},r_{j}} x^{j}.
\nonumber
\end{align}
From initial conditions, we have
$ A^{(l_{\Sigma},r_{j})}(1)
  =
 -\ep\epbar G_{j}(1)$
and
$
 C_{l_{\Sigma},r_{j}}
  =
 0$.
Hence we obtain
\begin{align}
 A^{(l_{\Sigma},r_{j})}
  =
 \ep\epbar 
   \sum_{i\in\mathcal{L}}\frac{\lambda_{i}}{i} (y^{i}-1)
      G_{j} \frac{x'}{x}
 -\epbar G_j x.
\nonumber
\end{align}
This leads to (\ref{lem2.2}) for $j\in\{2,3,\dots,\rmax-1\}$.

Note that 
$ A^{(l_{\Sigma},r_1)}
 = 
 A^{(l_{\Sigma},r_{\Sigma})}
-\sum_{j=2}^{\rmax-1} A^{(l_{\Sigma},r_{j})}$.
We have
\begin{align}
 A^{(l_{\Sigma},r_{1})}
  =
 \ep\epbar 
   \sum_{i\in\mathcal{L}}\frac{\lambda_{i}}{i} (y^{i}-1)
      \bigl(G_{1} \frac{x'}{x} -1\bigr)
 -\epbar G_1 x.
\nonumber
\end{align}
Hence we obtain (\ref{lem2.2}).

\paragraph{Proof of (\ref{lem2.3})}
Since (\ref{diff_lrSs}) is a first order differential equation,
it can be solve as follows:
\begin{IEEEeqnarray}{RL}
 S^{(l_k,l_s;r_{\Sigma})}
  =&
  x^{\rmax}\int\frac{1}{x^{\rmax}}
  \bigl[
    \frac{ D^{(l_k, r_{\Sigma})} }{k\E{l}_k}
   -\frac{ D^{(l_s, r_{\Sigma})} }{s\E{l}_s}
     \nonumber \\
 &\hspace{16mm}-(\rmax-1)\frac{x'}{x}S^{(l_k,l_s;l_{\Sigma})}
  \bigr] \diff{y}
 +Cx^{\rmax}.
 \nonumber
\end{IEEEeqnarray}
Note that
\begin{IEEEeqnarray}{L}
  \frac{ D^{(l_k, r_{\Sigma})} }{k\E{l}_k}
 -\frac{ D^{(l_s, r_{\Sigma})} }{s\E{l}_s}
 -(\rmax-1)\frac{x'}{x}S^{(l_k,l_s;l_{\Sigma})}
  \nonumber \\
 \hspace{5mm}=
  K^{(l_k,r_{\Sigma})} 
 -K^{(l_s,r_{\Sigma})} ,
 \nonumber
\end{IEEEeqnarray}
where
\begin{IEEEeqnarray}{L}
 K^{(l_k,r_{\Sigma})} 
  \nonumber \\
  \hspace{3mm}:=
 \ep G_{\Sigma}
  \bigl[
   -2\frac{x'}{e}y^{k}
   +(k-1)y^{k-2}
   +\frac{2(x')^{2}-x''x}{x^{2}}\frac{y^{k}-1}{k}
  \bigr]
  \nonumber \\
 \hspace{8mm}+(\rmax-1)\frac{x'}{x}\ep
  \bigl(
    y^{k}
   -\frac{x'y+x}{x}\frac{y^{k}-1}{k}
  \bigr).
 \nonumber
\end{IEEEeqnarray}
Note that
\begin{align}
 &x^{\rmax}\int\frac{1}{x^{\rmax}}  K^{(l_k,r_{\Sigma})} \diff{y}
=
 -\ep \bigl( G_{\Sigma}\frac{x'}{x} - 1 \bigr) \frac{y^{k}-1}{k} 
 +\ep G_{\Sigma}    y^{k-1}.
 \nonumber
\end{align}
Thus we have
\begin{IEEEeqnarray}{RL}
  S^{(l_k,l_s;r_{\Sigma})}
   =&
 -\ep \bigl( G_{\Sigma}\frac{x'}{x} - 1 \bigr) 
   \bigl(
     \frac{y^{k}-1}{k} 
    -\frac{y^{s}-1}{s} 
   \bigr)
  \nonumber \\
 &{}+\ep G_{\Sigma}    (y^{k-1} -y^{s-1})
 +C x^{\rmax}.
\nonumber
\end{IEEEeqnarray}
From the initial covariance, we have
$ S^{(l_i,l_s;r_{\Sigma})}(1) = 0 $
and
$ C = 0$.
This leads to (\ref{lem2.3}).
\paragraph{Proof of (\ref{lem2.4})}
In a way similar to Section \ref{sec_der_lem2.2},
we can obtain (\ref{lem2.4}).
\end{IEEEproof}
\subsubsection{Proof of (\ref{delt_lr})}

From definitions of $S^{(l_k,l_s;r_j)}$ and $A^{(l_{\Sigma},r_j)}$,
we see that
\begin{IEEEeqnarray}{RCL}
 \delt{l_s}{r_j}
  &=&
  \frac{s\E{l}_s}{e}
   \Bigl(
     A^{(l_{\Sigma},r_{j})}
   - \sum_{k\in\mathcal{L}}\E{l}_k S^{(l_k,l_s;r_j)}
   \Bigr)
\nonumber \\
  &=&
  \bigl[ F\frac{s\E{l}_s}{e} - \ep \E{l}_s (y^{s}-1) \bigr]
  \bigl( \frac{x'}{x}G_{j} - I_{\{j=1\}} \bigr)
   \nonumber \\
  &&-\frac{s\E{l}_s}{e}G_{j}
  (\frac{F'+x}{2} -\ep x y^{s}).
\nonumber
\end{IEEEeqnarray}
Thus, we obtain (\ref{delt_lr}).

\subsection{Lemma and Proof of (\ref{delt_rr})}\label{proof_delt_rr}
In this section, we introduce a lemma to prove (\ref{delt_rr})
and we prove (\ref{delt_rr}).
\subsubsection{Lemma to Prove (\ref{delt_rr})}
\begin{lemma} \label{lemma3} \upshape
 We define 
 $\delt{r_j}{r_{\Sigma}} 
  := 
  \sum_{k\in\bar{\calR}}\delt{r_j}{r_k}$ 
 and 
 $\delt{r_{\Sigma}}{r_\Sigma} 
  := 
  \sum_{j\in\bar{\calR}}\delt{r_j}{r_{\Sigma}}$.
For $j\in\bar{\calR}$, we have the following equations.
 \begin{IEEEeqnarray}{RCL}
  \delt{r_{\Sigma}}{r_{\Sigma}}
   &=&
  -F\bigl( \frac{x'}{x}G_{\Sigma}-1\bigr)^{2}
  +F' G_{\Sigma}\bigl(\frac{x'}{x} G_{\Sigma} -1\bigr) 
   \nonumber \\
  &&-G_{\Sigma}^{2} \sum_{i\in\calL} i\ep^{2} \lambda_{i} y^{2i-2}
  +\rmax^{2}\E{r}_{\rmax}^{2}
  -V_{\rmax,\rmax},
   \label{lem3.1}  \\
  \delt{r_{j}}{r_{\Sigma}}
   &=&
  -F\bigl( \frac{x'}{x}G_{\Sigma} - 1 \bigr)
    \bigl( \frac{x'}{x}G_{j} - I_{\{j=1\}} \bigr)
   \nonumber \\
  &&+ F' G_{j} \bigl( \frac{x'}{x}G_{\Sigma} -1 \bigr)
  -G_{\Sigma}G_{j}\sum_{i\in\calL}\ep^{2}i \lambda_{i} y^{2i-2}
  \nonumber \\
  &&+\rmax\E{r}_{\rmax}x G_{j}
  +V_{j,\rmax}
  +\frac{F'-x}{2}
   \bigl(
     G_{j}    
    -I_{\{j=1\}}G_{\Sigma} 
   \bigr)
   \nonumber \\
 &&+I_{\{j=1\}}\rmax\E{r}_{\rmax}(e-x).
  \label{lem3.2}
 \end{IEEEeqnarray}
\end{lemma}
We use (\ref{lem3.1}) to prove of the basis for the mathematical
induction in proof of (\ref{lem3.2}).
Similarly, we use (\ref{lem3.2}) to prove of the basis for the
mathematical induction in proof of (\ref{delt_rr}).

\begin{IEEEproof}
First, we derive differential equations.
We define 
$\delt{r_{i}}{r_{\Sigma}}
 :=
 \sum_{j\in\bar{\calR}} \delt{r_i}{r_j}$,
$\delt{r_{\Sigma}}{r_{\Sigma}}
 :=
 \sum_{i\in\bar{\calR}} \delt{r_i}{r_{\Sigma}}$
and
$\delt{r_{\rmax}}{r_{j}} 
  :=
 \delt{l_{\Sigma}}{r_j} - \delt{r_{\Sigma}}{r_j}$
.
From covariance evolution (\ref{CE}), we get
\begin{IEEEeqnarray}{RCL}
 \divi{\delt{r_i}{r_j}}{y}
  &=&
 -\frac{x'}{x}
    \bigl[
       i\delt{r_{i+1}}{r_j}
      +j\delt{r_{j+1}}{r_i}
      -(i+j)\delt{r_j}{r_i}
    \bigr]
   \nonumber \\
   &&+D^{(r_i,r_j)},
 \label{diff_rr}
\end{IEEEeqnarray}
where
\begin{IEEEeqnarray}{RCL}
 D^{(r_i,r_j)}
  &:=&
  \sum_{k\in\mathcal{L}}\frac{2a-k-1}{y^{2}}
   \bigl(
     \delt{l_{k}}{r_j}G_{i}
    +\delt{l_{k}}{r_i}G_{j}
   \bigr)
  \nonumber \\
 &&-x\hat{f}^{(r_i,r_j)}.
\nonumber
\end{IEEEeqnarray}
Define 
$ D^{(r_i,r_{\Sigma})}
 :=
  \sum_{j\in\calL}D^{(r_i,r_j)}$
and
$ D^{(r_{\Sigma},r_{\Sigma})}
 :=
  \sum_{i\in\calL}D^{(r_i,r_{\Sigma})}$.
For $\delt{r_i}{r_{\Sigma}}$, we have
\begin{IEEEeqnarray}{RCL}
 \divi{\delt{r_i}{r_{\Sigma}}}{y}
  &=&
 -\frac{x'}{x}
   \bigl[
     i\delt{r_{i+1}}{r_{\Sigma}}
    -(\rmax+i)\delt{r_i}{r_{\Sigma}}
   \bigr]
   \nonumber \\
 &&-\frac{x'}{x}  (\rmax-1)\delt{l_{\Sigma}}{r_i} 
 +D^{(r_i,r_{\Sigma})}.
\nonumber
\end{IEEEeqnarray}
The sum over this equation for $i\in\bar{\calR}$ is written as follows:
\begin{IEEEeqnarray}{RL}
 \divi{\delt{r_{\Sigma}}{r_{\Sigma}}}{y}
  =&
   -2\frac{x'}{x}
      \bigl[(\rmax-1)\delt{l_{\Sigma}}{r_{\Sigma}} 
           -\rmax\delt{r_{\Sigma}}{r_{\Sigma}} \bigr]
   \nonumber \\
  &+D^{(r_{\Sigma},r_{\Sigma})}.
 \label{diff_rSS}
\end{IEEEeqnarray}

\paragraph{Proof of (\ref{lem3.1})}
Since (\ref{diff_rSS}) is a first order differential equation,
it can be solve as follows:
\begin{IEEEeqnarray}{RCL}
 \delt{r_{\Sigma}}{r_{\Sigma}}
  &=&
 x^{2\rmax} \int \frac{1}{x^{2\rmax}} 
  \bigl[
    D^{(r_{\Sigma},r_{\Sigma})}
   -2(\rmax-1)\frac{x'}{x}\delt{l_{\Sigma}}{r_{\Sigma}}
  \bigr]
   \diff{y}
  \nonumber \\
 &&+x^{2\rmax} C_{r_{\Sigma},r_{\Sigma}}
 \nonumber \\
  &=&
  -F\bigl( \frac{x'}{x}G_{\Sigma}-1\bigr)^{2}
  +G_{\Sigma}\bigl(\frac{x'}{x} G_{\Sigma} -1\bigr) F' 
   \nonumber \\
  &&-G_{\Sigma}^{2} \sum_{i\in\calL} i\ep^{2} \lambda_{i} y^{2i-2}
  +C_{r_{\Sigma},r_{\Sigma}} x^{2\rmax} ,
\nonumber
\end{IEEEeqnarray}
with a constant $C_{r_{\Sigma},r_{\Sigma}}$ which can be determined from 
initial conditions.
From initial conditions, we have
\begin{IEEEeqnarray}{RCL}
  \delt{r_{\Sigma}}{r_{\Sigma}}(1)
 &=&
 \lambda'(1)\ep\epbar (\rmax\rho_{\rmax} \ep^{\rmax-1}-1)^{2}
+\ep\epbar
  \nonumber \\
&&-2\ep\epbar \rmax \rho_{\rmax} \ep^{\rmax-1}
 +\rmax\rho_{\rmax}\ep^{\rmax}
-\rmax\rho_{\rmax}\ep^{2\rmax}
 \nonumber
\end{IEEEeqnarray}
and
$ C_{r_{\Sigma},r_{\Sigma}}
  =
 \rmax^{2}\rho_{\rmax}^{2}
-\rmax\rho_{\rmax}
$.
Thus we have
\begin{IEEEeqnarray}{RCL}
  \delt{r_{\Sigma}}{r_{\Sigma}}
   &=&
  -F\bigl( \frac{x'}{x}G_{\Sigma}-1\bigr)^{2}
  +G_{\Sigma}\bigl(\frac{x'}{x} G_{\Sigma} -1\bigr) F' 
   \nonumber \\
  &&-G_{\Sigma}^{2} \sum_{i\in\calL} i\ep^{2} \lambda_{i} y^{2i-2}
  +\rmax^{2}r_{\rmax}^{2}
  -V_{\rmax,\rmax}.
 \nonumber
\end{IEEEeqnarray}
This leads to (\ref{lem3.1}).
\paragraph{Proof of (\ref{lem3.2})}
In a way similar to Section \ref{sec_der_lem2.2},
we can obtain (\ref{lem3.2}).
\end{IEEEproof}

\subsubsection{Proof of (\ref{delt_rr})}
(\ref{diff_rr}) can be solve as follows:
\begin{IEEEeqnarray}{L}
 \delt{r_i}{r_j}
  \nonumber \\
  =
 x^{i+j}
 \int \frac{1}{x^{i+j}}
  \bigl(
    D^{(r_i,r_j)}
   -\frac{x'}{x}i\delt{r_{i+1}}{r_j}
   -\frac{x'}{x}j\delt{r_{i}}{r_{j+1}}
  \bigr) \diff{y}
  \nonumber \\
 \hspace{5mm}+C_{r_i,r_j}x^{i+j}.
 \label{int_rr}
\end{IEEEeqnarray}
This equation can be solved by mathematical induction
for $i,j \in\{2,3,\dots,\rmax-1\}$.
Note that from (\ref{lem3.2})
\begin{IEEEeqnarray}{L}
 \delt{r_j}{r_{\rmax}}
  \nonumber \\
  \hspace{5mm}=
  G_j G_{\rmax}
   \bigl[
    -F \bigl(\frac{x'}{x}\bigr)^{2}
    +F'\frac{x'}{x} 
    -\sum_{s\in\calL}\ep^{2}s\lambda_s y^{2s-2}
    +x^{2}
   \bigr]
  \nonumber \\
 \hspace{10mm}-V_{j,\rmax},
 \nonumber
\end{IEEEeqnarray}
for $j\in\{2,3,\dots,\rmax-1\}$.
Using the same method in the induction step, we see that
$\delt{r_{\rmax-1}}{r_{\rmax-1}}$ fulfill (\ref{delt_rr}).

We show that if 
$\{\delt{r_i}{r_j} \mid i,j\in\{2,3,\dots,\rmax-1\}, i+j = k+1\}$
fulfill (\ref{delt_rr}), 
then
$\{\delt{r_i}{r_j} \mid i,j\in\{2,3,\dots,\rmax-1\}, i+j = k\}$
fulfill (\ref{delt_rr}).
Using the induction hypothesis, we can solve (\ref{int_rr})
\begin{IEEEeqnarray}{L}
 \delt{r_i}{r_j}
  \nonumber \\
  =
  S_i S_j 
   \bigl[
    -F\bigl(\frac{x'}{x}\bigr)^{2}
    +F'\frac{x'}{x} 
    +x^{2} 
   -\sum_{s}\ep^{2}s\lambda_s y^{2s-2}
   \bigr] 
 -V_{i,j}
   \nonumber \\
 \hspace{5mm}+I_{\{i=j\}} i\sum_{s}\rho_{s} \binom{s-1}{i-1}
   \bigl[
     x^{i}\xbar^{s-i}
    -\binom{s-i}{i}x^{i}(-x)^{i}
   \bigr]
   \nonumber \\
 \hspace{5mm} +C_{r_i,r_j}x^{i+j}.
 \nonumber
\end{IEEEeqnarray}
From the initial condition, we get
\begin{IEEEeqnarray}{rCl}
 C_{r_i,r_j} 
  =
 I_{\{i=j\}}i\sum_{s}\rho_{s}\binom{s-1}{i-1}\binom{s-i}{i}(-1)^{i}.
 \nonumber
\end{IEEEeqnarray}
Thus, we have (\ref{delt_rr}) for $i,j \in\{2,\dots,\rmax-1\}$.

Note that 
$\delt{r_i}{r_1} 
  = 
 \delt{r_i}{r_{\Sigma}} - \sum_{j=2}^{\rmax-1}\delt{r_i}{r_j}$.
We show that $\delt{r_i}{r_1}$ fulfill (\ref{delt_rr}) 
for $i\in\bar{\calR}$.
Hence we obtain  (\ref{delt_rr}).

\section{Conclusion}
In this paper, we have analytically solved the covariance evolution
for irregular LDPC code ensembles.
We have also obtained the slope scaling parameter.




\ifCLASSOPTIONcaptionsoff
  \newpage
\fi


\end{document}